\documentclass[11pt]{article}

\usepackage{epsf}

\newcommand{\beq}{\begin{equation}}
\newcommand{\eeq}{\end{equation}}
\newcommand{\beqa}{\begin{eqnarray}}
\newcommand{\eeqa}{\end{eqnarray}}
\newcommand{\tr}{\mbox{tr}}
\newcommand{\pexp}{\mbox{Pexp}}
\newcommand{\fot}{\frac{1}{2}}
\newcommand{\fof}{\frac{1}{4}}
\def\a{\alpha}
\def\b{\beta}
\def\c{\gamma}
\def\d{\delta}
\def\e{\epsilon}
\def\w{\omega}

\def\intx{\int\! d^3x\,}
\def\inty{\int\! d^3y\,}
\def\intz{\int\! d^3z\,}
\def\ints{\int\! ds\,}
\def\intt{\int\! dt\,}
\def\intu{\int\! du\,}
\def\intv{\int\! dv\,}
\def\intA{\int\! {\cal D} A \,}

\begin{document}

\begin{flushright}
gr-qc/9512036
\\
December 20, 1995
\end{flushright}

\vfill

\begin{center}{\Large\bf
On the constraint algebra of quantum gravity \\[4mm] 
in the loop representation
}
\end{center}
\vfill

\begin{center}
\large Bernd Br\"ugmann
\end{center}

\begin{center}
{\em Max-Planck-Institut f\"ur Gravitationsphysik,
 \\ Schlaatzweg 1, 14473 Potsdam, Germany \\}
{\tt bruegman@aei-potsdam.mpg.de}
\end{center}

\vfill

\begin{center}
\large\bf Abstract
\end{center}
\medskip

\noindent
Although an important issue in canonical quantization, the problem of
representing the constraint algebra in the loop representation of quantum
gravity has received little attention.  The only explicit computation was
performed by Gambini, Garat and Pullin for a formal point-splitting
regularization of the diffeomorphism and Hamiltonian constraints. It is
shown that the calculation of the algebra simplifies considerably when the
constraints are expressed not in terms of generic area derivatives but
rather as the specific shift operators that reflect the geometric meaning
of the constraints.

\vspace*{\fill}
\newpage

\section{Introduction}

A key problem of canonical quantization is the incorporation of the
classical constraints. In general, we have to expect that there are
many different and in their outcome genuinely inequivalent ways to
import classical symmetries into a quantum theory (for a recent
example, see \cite{Ja}), and a famous example for the resolution of
such ambiguities is the critical dimension of string theory. 
How to define and how to regulate the constraint operators is directly
related to the representation of the constraint algebra in the quantum
theory. Typically, the constraint operators are not uniquely determined by
the classical constraints, and demanding well-definedness of the quantum
constraint algebra is an important restriction on the choice of
representation.

Here we study the constraint algebra of vacuum general relativity in
3+1 dimensions in the framework of Dirac quantization, in which the
classical constraints $C$ are elevated to quantum operators $\hat C$
and imposed as operator relations on the wavefunctions, $\hat C \psi =
0$. Given a definition of the constraint operators, the question is
whether there are anomalies in the constraint algebra, and if so, how
to treat them. Currently there are no rigorous results about the
constraint algebra of quantum gravity because of difficulties to
define the constraint operators, e.g.\ issues of factor ordering and
regularization arise.

In the canonical formulation of vacuum
general relativity there are two types of constraints, the
three-dimensional diffeomorphism constraint $D$ and the Hamiltonian
constraint $H$, which satisfy the following algebra:
\beqa 
\{ D(v), D(w) \} & = & D({\cal L}_v w),
\label{dd}
\\
\{ D(v), H(N) \} & = & H({\cal L}_v N),
\label{dh}
\\
\{ H(M), H(N) \} & = & D(g^{ab} \w_b), 
\label{hh}
\eeqa 
where
$\w_b = M \partial_b N - N \partial_b M$,
$g^{ab}$ is the inverse three-metric, $v$ and $w$ are vector fields,
and $M$ and $N$ are scalar densities of weight -1, all on a three manifold
which we choose to be compact. 

The constraint algebra of general relativity has two important features.
First, the algebra closes but only for structure functions containing one
of the geometric variables in (\ref{hh}). And second, neither $D$ nor $H$
form an ideal, and therefore one cannot construct a reduced phase space
with respect to just one of the constraints. For example, the
diffeomorphism constraint would form an ideal if the constraint on the
right-side of (\ref{dh}) was $D$, but as it stands, there are no invariant
subspaces that are necessary for phase space reduction.

For canonical quantization of general relativity a very fruitful
approach has turned out to be the Rovelli-Smolin loop representation
\cite{RoSm:lrep} 
based on the Ashtekar variables \cite{As,As:book}. For a motivation of
the loop representation and a discussion of its strong and weak
points, see for example \cite{Br:review}.  In this paper we focus on
the technical problem of how to construct the quantum constraint
operators in the loop representation, and present a formal calculation
of the constraint algebra in the loop representation.
 
The loop representation, in which states are functionals of loops, can
be obtained from the connection representation, in which states are
functionals of the Ashtekar connection, through the so-called loop
transform.  We introduce a for our purposes particularly convenient
form of the constraints via the transform.  The level of our
discussion remains on the same heuristic level as the original
attempts to define the constraints \cite{RoSm:lrep,Bl,Ga,BrPu}. In
particular, the loop transform is not defined rigorously.  Although
there now does exist a rigorous definition of the two representations
and the loop transform based on distributional connections
\cite{AsLeMaMoTh}, which brings about certain changes to the whole
formalism, it is not yet known how to treat the Hamiltonian constraint
rigorously.  Note that the constraint operators in the loop
representation can also be obtained without the loop transform
directly from the Wilson-loop variables, with the same result
\cite{Br:thesis}.

The definition of the constraint operators requires a regularization,
and we use a point-splitting regularization. The problem with
point-splitting is that it introduces a background dependence which
breaks diffeomorphism invariance and which survives in the limit that
the regulators are removed. Further problems are that the constraint
operators should act on a Hilbert space of states, but we do not have
an inner product, and that the reality conditions of the Ashtekar
formulation have not been implemented, and the constraint
algebra of complex relativity might differ from that of real relativity. 
There have been important advances on these three issues
(e.g. \cite{RoSm:prl}, \cite{AsLe}, and \cite{reality},
respectively), but for the above reasons all calculations of the
constraint algebra must still be called formal. What can be explored,
however, is what the current framework predicts for the constraint
algebra, and in the loop representation this is not a
trivial task.

One can distinguish four approaches to the constraint algebra in
the loop representation:
\begin{enumerate}

\item 
  Recall that in the connection representation, the constraint
  algebra, formally regulated by point-splitting, does close
  \cite{As:book}, and one can hope that this feature is preserved
  by the loop transform (which is as we already pointed out still
  problematic for the Hamiltonian constraint). This justifies the
  route most commonly taken until quite recently, namely to postpone
  the treatment of the constraint algebra in the loop
  representation. Given the definition of the constraint operators in
  the loop representation, one proceeds to study their kernel, to look
  for an inner product, defines observables, etc.\ without taking the
  constraint algebra into account.

\item 
  Given the definition of the constraint operators in the loop
  representation, the constraint algebra is computed explicitly
  \cite{GaGaPu}. The technical difficulties encountered explain
  strategy 1. One problem is that the differential operator on the
  space of loop functionals that appears in the Hamiltonian constraint
  \cite{Bl,Ga,BrPu}, the area derivative, is not very well studied,
  but now the machinery is available \cite{GaPu:book} that makes
  \cite{GaGaPu} possible.

\item One attempts a two-step procedure, first solving the diffeomorphism
  constraint, and then defining the Hamiltonian constraint operator
  only on the diffeomorphism invariant states.  This is how sometimes
  the original Rovelli-Smolin loop representation \cite{RoSm:lrep} is
  interpreted, namely as a theory of operators on knot invariants. For
  example, in the rigorous approach based on measures on the space of
  connections modulo gauge \cite{AsLe,Ba}, the measures are
  constructed to be diffeomorphism invariant first, and one later
  attempts to construct a ``diffeomorphism invariant'' Hamiltonian
  constraint operator. An analogous approach is taken in Loll's
  lattice gravity, which has led to important insights for
  certain geometric operators \cite{Lo}.

{}From the perspective of the quantum constraint algebra corresponding
to (\ref{dd})--(\ref{hh}), we have to observe that one cannot simply
first impose the diffeomorphism constraint without consequences for
the Hamiltonian constraint, since as mentioned above the constraints
are intertwined such that there are no ideals. Let us give a
simple-minded example. Suppose
that $D(v)\psi = 0$ and that $H(N)$ does not map states out of the
kernel of $D(v)$, i.e.\ $D(v) H(N) \psi = 0$. Then $[D(v),H(N)] = 0$,
and by (\ref{dh}), $H({\cal L}_vN) = 0$, which since it holds for all
$v$ and $N$ implies $H(N) = 0$ \cite{Re}. That is, if we want to work
exclusively on the space of solutions to the diffeomorphism
constraint, we have to consider the subspace that is invariant under
the Hamiltonian constraint, and then the constraint algebra implies
that this subspace must be the simultaneous kernel of the constraints.

If we want to define the symmetries of quantum gravity with the help
of the operator constraint algebra, we therefore have to implement the
constraint algebra before imposing the diffeomorphism constraint.

In this context it is perhaps worth pointing out that one of the main
attractive features of the loop representation, namely that a few
non-trivial (formal) solutions to both constraints are known, is not
derived from a two-step implementation of the constraints. Rather, one
attempts to find the intersection of the kernels of the constraints on
the unreduced space of loop functionals \cite{solutions}.

\item One introduces (almost trivial) matter variables that allow the
  definition of a preferred time-slicing. The Hamiltonian constraint
  operator then turns into a proper Hamiltonian operator, and the
  Wheeler-DeWitt equation into a proper Schr\"odinger equation
  \cite{RoSm:prl}. The main advantage is that the regularized
  Hamiltonian operator is diffeomorphism invariant, which opens up a
  whole range of interesting topics in diffeomorphism invariant
  dynamics. The problem of representing the constraint algebra is
  easily solved because only the diffeomorphism algebra remains to be
  represented, the information about the Hamiltonian constraint is now
  contained in the Schr\"odinger equation. In this sense, a preferred
  time-slicing also defines a two-step procedure.  One should
  remember, and emphasize, however, that the construction of matter
  clocks works only {\it locally}.  The resulting theory is therefore
  only an approximation to what is usually called full quantum
  gravity. It would be very interesting to gain some control over the
  approximation, for example in reduced models.  Even so, the prospect
  of diffeomorphism invariant dynamics is very interesting, and may be
  physically quite relevant.

\end{enumerate}

To summarize, one can either try to implement the classical
constraints directly, 1 and 2, or to give a special treatment to the
Hamiltonian constraint, 3 and 4. For the latter, there are good
motivations, like the observation that while the diffeomorphism
constraint is linear in the momenta and therefore generates a type of
gauge symmetry, the Hamiltonian constraint is quadratic in the momenta
and does not possess simple gauge orbits. Also, to obtain the
conventional interpretation of time in quantum gravity a special
treatment of the Hamiltonian constraint may be necessary.

Here we do not choose to treat the constraints differently, but want
to explore as in 2 whether the classical constraint algebra has a
complete representation in the loop representation of canonical
quantum gravity. To state the outcome of the calculation, the
resulting algebra maintains the structure of the classical algebra
with a particular choice of factor ordering for $D(g^{ab}\w_b)$, which
becomes necessary because of the presence of the metric. The level of
rigor is equivalent to that in the connection representation, and upon
removal of the point-splitting there are no anomalies. Let us point
out that the formal nature of the point-splitting regularization does
not allow us to decide whether there actually are anomalies in the
constraint algebra of quantum gravity.

Our contribution is to show that starting with a different form of the
constraints the result of \cite{GaGaPu} can be obtained in a simpler
manner.  The simplification comes about by the observation that the
constraints are not just based on generic area derivatives, but rather
on more specialized geometric operators, certain shift operators.
This makes the algebra managable to the extent that the
point-splitting regularization can now be studied further along the
lines of \cite{Bo}, where the removal of the regulators is examined in
detail for the constraint algebra in the connection representation.
In terms of shift operators, the constraint operators may
in fact be compatible with the rigorous regularization techniques
coming from distributional connections, with which it is particularly
hard to represent the field strength of the Ashtekar connection that
directly corresponds to the area derivative via the loop transform.

The paper is organized as follows. In section 2, we define the
constraints in the connection and the loop representation and discuss
the point-splitting regularization. In section 3, we introduce the
basic loop derivative commutators. Section 4 contains the calculation
of the constraint algebra in the loop representation. In section 5, we
conclude with a few comments.


\section{The constraints and their representation}

The constraints of quantum gravity in the loop representation have
been derived in various ways \cite{RoSm:lrep,Bl,Ga,BrPu} with
essentially the same result \cite{Br:thesis}. Since our method to compute the
constraint algebra depends crucially on the form of the constraint
operators, let us give a brief derivation via the loop transform.

The Ashtekar variables are a connection $A_a^i(x)$ and a vector
density $E^{ai}(x)$ of weight one, both complex, on a compact
three-manifold $\Sigma$. Tangent space indices are denoted by
$a,b,\ldots$, internal indices are denoted by $i,j,\ldots$. The
internal gauge group is $SU(2)$, and following \cite{RoSm:lrep} we
choose generators $\tau^i$ such that
\beq 
[\tau^i, \tau^j] = \epsilon^{ijk} \tau^k.
\eeq
The algebra-valued variables are obtained by contraction, e.g.\ $A_a =
A_a^i \tau^i$.
The inverse metric is given by
\beq
g g^{ab} = E^{ai} E^{bi}, 
\eeq
where $g$ is the determinant of $g_{ab}$ (insuring the correct density
weight).

The constraints of general relativity which satisfy the algebra
(\ref{dd})--(\ref{hh}) are
\beqa
D(v) &=& \intx v^a E^{bi} F^i_{ab} + G(v),
\\
H(N) &=& \intx N \epsilon^{ijk} E^{ai} E^{bj} F^k_{ab}.
\eeqa
The vector constraint $\intx v^a E^{bi} F^i_{ab}$ generates
diffeomorphisms up to an internal gauge transformation, which is
compensated by the term $G(v)$, and a gauge depending term also
appears in (\ref{hh}). (The sign convention for the Poisson
brackets in (\ref{dd})--(\ref{hh}) is opposite to that of (\cite{As:book}).)

In the connection representation, wave functions are functionals of the
Ashtekar connection, $\psi[A]$, and the operators corresponding to the
connection and the triad are represented by
\beq
     \hat A_a^i(x) = A_a^i(x), \quad \hat E^{ai}(x) = -
     \frac{\delta}{\delta A_a^i(x)},
\eeq
where $\hbar = 1$ and a complex $i$ has been absorbed in the definition of
$\hat E^{ai}$.

The first non-trivial issue we have to face is regularization. While the
type of point-splitting that we use has been commonly applied in many
places, let us proceed slowly since there are different
prescriptions for in which order the various regulators have to be
removed.

The necessity for regularization arises at this point because the
metric and the constraints are products of operators at the same
point. We introduce a point-splitting based on a background metric and
a regulator $f_\epsilon(x,y)$ satisfying
\beq
\lim_{\epsilon\rightarrow 0} f_\epsilon(x,y) = \delta^3(x,y).
\eeq
For the calculations that follow we fix a coordinate system and
require that $f_\e(x,y)$ is a smooth function of $x-y$.

As discussed in \cite{TsWo,FrJa}, which refer explicitly to the
constraint algebra, a `full' point-splitting should be applied, that
is, all points that appear in an operator product should be split.
Also see \cite{BrGaPu:jones}, where it is shown that only in a
particular factor-ordering the connection representation is related to
the loop representation, and that the vector constraint in this factor
ordering only gives rise to diffeomorphisms in the connection
representation if a symmetric point-splitting, $f_\epsilon(x,y) =
f_\epsilon(y,x)$ is employed.  Following these considerations we
define the operators
\beqa
g^{ab}_{\e\e'}(x) &=& \inty \intz f_\e(x,y) f_{\e'}(x,z) 
\frac{\d}{\d A_a^i(y)} 
\frac{\d}{\d A_b^i(z)}, 
\\
D_\e(v) &=&
- \intx \inty f_\e(x,y) v^a(x) \frac{\d}{\d A_b^i(x)} F^i_{ab}(y) + G(v),
\\ 
H_{\e\e'}(N) &=&
\intx \inty \intz f_\e(x,y) f_{\e'}(x,z) N(x) \e^{ijk}
\frac{\d}{\d A_a^i(y)}
\frac{\d}{\d A_b^j(z)}
 F^k_{ab}(x).
\eeqa

For the particular calculations that follow, some of the regulators
can be removed since there do not arise
singularities that require them.  (As just recalled, this is not true
for the vector constraint algebra in the connection representation.)

The transition to the loop representation is made via a formal transform,
the loop transform
\beq
\psi[\eta] = \intA \tr U_\eta \psi[A],
\eeq
where $U_\eta$ is the holonomy matrix of $A_a^i$ around the loop $\eta$,
\beq
U_\eta = \pexp \ints \dot\eta^a(s) A_a(\eta(s)).
\eeq

In order to transfer the operators that we are interested in in the
connection representation, $O_C$, to operators in the loop
representation, $O_L$, one performs a formal
partial integration for any occurrence of a functional derivative with
respect to $A_a^i$, so that 
\beq
O_L \psi[\eta] \equiv \intA \tr U_\eta O_C \psi[A] 
=  \intA  (O^+_C\tr U_\eta) \psi[A].
\label{parint}
\eeq
All that is needed for an explicit definition of $O_L$ as loop
operator is a transfer relation of the operators on the Wilson
loops that expresses the operation on the connection dependence of the
Wilson loop purely as an operation on its loop dependence, 
\beq
O_C^+ \tr U_\eta = O_L \tr U_\eta.
\eeq

This construction is directly applicable to the metric and the
constraints.  The Wilson loop satisfies
\beqa
\frac{\d}{\d A_a^i(x)} \tr U_\eta &=& 
\ints \d^3(x, \eta(s)) \dot\eta^a(s) \tr (U_{ss} \tau_i ),
\label{dA}
\\
\frac{\d}{\d \eta^a(s)} \tr U_\eta &=& 
\dot\eta^b(s) F^i_{ab}(\eta(s)) \tr (U_{ss} \tau_i ),
\label{transfer}
\eeqa
where $U_{ss}$ denotes the holonomy from $\eta(s)$ once around the
loop.  Variations of the Wilson loop with respect to the connection
and the loop are not completely unrelated, and for the operators under
considerations there exist transfer relations precisely because of
that. In fact, (\ref{transfer}) is the one transfer relation we need,
relating the loop derivative on the loop representation side with the
multiplication and insertion of a field strength on the connection
representation side.

At this point it turns out to be quite advantageous to introduce a new
piece of notation. Let $T_s^i$ be the loop operator that inserts a
generator at parameter value $s$ into the holonomy,
\beq
T_s^i \tr U = \tr U_{ss} \tau^i.
\eeq
Partial integration as in (\ref{parint}) and evaluating the
functional derivative with respect to the connection as in (\ref{dA}) leads
to 
\beqa
D(v) \psi[\eta] &=&
\intA \left(\ints v^a(\eta(s))  \dot\eta^b(s)
F^i_{ab}(\eta(s)) T^i_s \tr U_\eta \right) \psi[A]
\\ 
H_{\e\e'}(N) \psi[\eta] &=&
\intA \left(
\intx\ints\intt f_\e(x,\eta(s)) f_\e(x,\eta(t)) N(x) 
\dot\eta^a(s) \dot\eta^b(t) \right.
\nonumber \\
&& \quad \left.
\e^{ijk} T_s^i T_t^j F^k_{ab}(x)
\tr U_\eta \right) \psi[A]
\label{Htransfer}
\\
g^{ab}_{\e\e'}(x) \psi[\eta] &=&
\intA \left(
\ints \intt f_\e(x,\eta(s)) f_{\e'}(x,\eta(t)) 
\dot\eta^a(s) \dot\eta^b(t) T_s^j T_t^j
\tr U_\eta \right) \psi[A].
\label{gtransfer}
\eeqa
We have removed the regulator in $D(v)$, and $G(v)$ does not contribute
since the Wilson loops are gauge invariant.

A single insertion operator could be transferred to the loop
representation by introducing functionals of loops with a marked
point, $T_s^i\psi[\eta]$. Such an extension is not necessary if
the insertion operator $T_s^i$ is combined with a field strength as in
$D(v)$, or if there are two insertions as in $g^{ab}$, or if
there are two insertions and a field strength as in $H(N)$.  The
double insertions that arise combine due to the trace identities for
$SL(2,C)$ matrices to the following rerouting operations,
\beqa
T_s^i T_t^i \tr U &=& \fof \tr U - \fot \tr U_{ts} \tr U_{st}
\\ 
\e^{ijk} T_s^i T_t^j \tr U &=& 
\fot (\tr U_{st} \tr U_{ts} \tau^k - 
      \tr U_{st} \tau^k \tr U_{ts})  
\eeqa
where $U_{st}$ denotes the parallel transport from $s$ to $t$ going
around the loop in the positive direction. That is, if $s > t$,
$U_{st} = U_{s1} U_{0t}$. Note that the resulting holonomies in
(\ref{Htransfer},\ref{gtransfer}), e.g.\
$\tr U_{st}$, in general refer to open paths, which however reduce to
closed loops in the limit that the regulator is removed.

As demonstrated, taking two derivatives with respect to $A^i_a$ leads
to the well-known rerouting of the loop at intersections in the loop
representation.  The rerouting in $g^{ab}$ and $H(N)$ is by no means a
trivial side-effect but actually crucial for the definition of the
operators and the closure of the constraint algebra. The Hamiltonian
constraint includes differentiation as well as rerouting. The
trace identities capture the fact that the internal gauge group is
complexified $SU(2)$ and not some other group.

The rerouting has to be denoted in some way, and we find it simplest
to not resolve the insertion operators into rerouted loops whenever
possible. For example it is then irrelevant whether $s < t$ or $t<s$,
and where the parameter origin of the loop is (as opposed to
\cite{GaGaPu}). This turns out to be an important technical point,
since the use of insertion operators allows us to separate the rerouting
operations from the other calculations and lead to a significant
simplification.

Whether we work with reroutings or insertion operators, we have
to deal with the case $s = t$. Note that 
for a given $s$, $\tr U_{ts} \tau^j U_{st} \tau^k$ is a 
function in $t$ with a finite step discontinuity at $t = s$.
The integrations in the metric and the Hamiltonian
constraint are evaluated by considering left- and right-sided
limits, $\ints\intt = \ints\int_0^{s^-}dt + \ints\int_{s^+}^1dt$. 

Note also that the product $T_s^i T_s^j \tr U$ has no
intrinsic meaning, since it is not clear where the second insertion
has to take place, before or after the first, but the left- and
right-sided limits $T_s^i T_{s^\pm}^j$ are well-defined.
Therefore, we define insertion operators on general loop functionals
that produce functionals of loops with a marked point, $T_s^i \psi[\eta]$,
and that inherit certain trace identities from the $\tau^i$. In
particular, 
\beqa
[T_s^i, T_t^j] &=& 0 \quad \mbox{for $s\neq t$},
\label{com1}
\\{}
[T_s^i, T_s^j] &:=& T_{s^-}^i T_{s^+}^j - T_{s^+}^i T_{s^-}^j
\label{com2}
\\{}
[T_s^i, T_s^j] &=& \e^{ijk} T^k_s.
\label{com2eps}
\eeqa


The transfer relation (\ref{transfer}) can be directly applied for
$D(v)$ and $g^{ab}_{\e\e'}(x)$. For $H(N)$ there remains the problem
that, as given in (\ref{Htransfer}), the field strength is evaluated
at $x$ and not on the loop as required in (\ref{transfer}). One can
either introduce at this point a more general type of loop derivative,
namely a path dependent area derivative (cmp.\ section 3.1), or use
that in the integrand
\beq F^i_{ab}(x) = F^i_{ab}(\eta(t)) + O(\e') \simeq F^i_{ab}(\eta(t)). 
\label{fxft}
\eeq 
All our calculations (and those of \cite{GaGaPu}) are performed only
to leading order in the point-splitting.

The transfer relation (\ref{transfer}), the commutator
(\ref{com2eps}), and attaching the field strength to the loop
(\ref{fxft}), allow us to arrive at our final form for the metric and
the constraints,
\beqa
D(v) &=&
\ints v^a(\eta(s)) \frac{\d}{\d \eta^a(s)},
\label{diffeo}
\\ 
H_{\e\e'}(N) &=&
\intx\ints\intt f_\e(x,\eta(s)) f_{\e'}(x,\eta(t)) \dot\eta^a(s)  N(\eta(t)) 
T_s^j [\frac{\d}{\d \eta^a(t)}, T_t^j],
\label{hamil}
\\
g^{ab}_{\e\e'}(x) &=& \ints \intt f_\e(x,\eta(s)) f_{\e'}(x,\eta(t)) 
\dot\eta^a(s) \dot\eta^b(t) T_s^j T_t^j,
\label{metric}
\eeqa
where in (\ref{hamil}) we have assigned the marked point property also
to the loop derivative.

The constraints in the loop representation that we have derived are
equivalent to the standard result \cite{Bl,Ga,BrPu}.  Apart from
notational differences for the rerouting, and order of $\e'$
differences in the regulators, there is, however, one very important
technical difference. The loop derivative that we introduce for the
transfer is not the area derivative, but just a special case thereof,
the ordinary functional derivative with respect to the loop, which has
the geometric interpretation of an infinitesimal shift operator.
Considering the other approaches, it is not clear why the functional
derivative should suffice, and we show below that for the removal of
the regulator one is forced to introduce the area derivative at
generic kinks of the loop.  In order to arrive at the above form we
had to take special care that the tangent vector to the loop is always
at the same point as the field strength as in (\ref{transfer}).



\section{Loop derivatives and some basic commutation relations}

Before moving on to the computation of the commutators of the
constraints, we take a more detailed look at the loop derivatives that
appear in the constraints, and analyze the basic commutators of the
derivatives and the rerouting operators.

\subsection{Commutators of loop derivatives for smooth loops}

The basic commutator for the computation of the constraint algebra is
that of two functional loop derivatives, 
\beq
[ \frac{\d}{\d \eta^a(s)}, \frac{\d}{\d \eta^b(t)} ] = 0.
\label{deldel}
\eeq
While this commutator vanishes, the commutator of two area derivatives
does not vanish, which is a good reason to attempt to rewrite the
Hamiltonian constraint in terms of functional derivatives. 
Let us discuss this important point in more detail.

The area derivative of a loop functional $\psi[\eta]$ is defined by
appending to $\eta$ an infinitesimal loop $\c^\d$ with area element
$\sigma^{ab}(\c^\d) = O(1/\d^2)$. In its general form, the area
derivative depends on a path $\pi_o^x$ (and its inverse $\pi_x^o$) from
the point $o$ at which all loops $\eta$ are supposed to be based to
the point $x$ where $\c^\d$ is attached. The definition of the path
dependent area derivative is
\beq
\Delta_{ab}(\pi_o^x) \psi[\eta] 
=
\lim_{\d\rightarrow0} \frac{ \psi[\pi_o^x \c^\d \pi_x^o \eta] - \psi[\eta]}
{\sigma^{ab}(\c^\d)},
\label{Delpath}
\eeq
where juxtaposition of loops denotes attachment of the loops at the
base point. From that definition follows the basic commutator used in
\cite{GaGaPu},
\beq
[ \Delta_{ab}(\pi_o^x) , \Delta_{cd}(\pi_o^y) ]
=
\Delta_{ab}(\pi_o^x)[\Delta_{cd}(\pi_o^y)],
\label{DelxDelx}
\eeq
where the brackets on the right-hand side indicate action on the path
dependence in $\pi_o^y$ only and not on the loop functionals. 
Note the peculiar form where the commutator of two derivatives is the
derivative of a derivative.

The area derivative that typically appears in the derivation of the
Hamiltonian constraint does not depend on arbitrary paths but on
portions of the loop on which the area derivative acts.  From
definition (\ref{Delpath}) with $\pi_o^x = \eta_o^x$, $x=\eta(s)$ we have a
definition for the parameter dependent area derivative,
\beq
\Delta_{ab}(s) \psi(\eta) := \Delta_{ab}(\eta_o^{\eta(s)}) \psi(\eta)
=
\lim_{\d\rightarrow 0} \frac{ \psi[\c^\d\circ_s \eta] - \psi[\eta]}
{\sigma^{ab}(\c^\d)}.
\label{DelsDelx}
\eeq

As remarked in \cite{GaPu:book}, the parameter dependent area
derivative is naturally more restricted in its applicability than the
generic, path dependent area derivative, but let us point out that
nevertheless it is all we need for the commutators of the
constraints. To be more specific, explicit calculation of the
infinitesimal loop variations show that $[\Delta_{ab}(s),
\Delta_{cd}(t)]$ is not expressible in terms of a parameter dependent
area derivative, but we also find that
\beq
\dot\eta^b(s) \dot\eta^d(t) [\Delta_{ab}(s), \Delta_{cd}(t)]  
=
- \delta(s,t) \frac{d}{ds} \Delta_{ac}(s).
\label{DelsDels}
\eeq 
One can peel of one of the tangent vectors obtaining a so-called
covariant loop derivative of an area derivative, but removing both
tangent vectors does not leave a single area derivative.

In pictures, parameter dependent area derivatives commute if the
infinitesimal loops are inserted at different parameters of the main
loop, fig.\ 1, but there is a nontrivial contribution if one of the
small loops is inserted onto the other, fig.\ 2. 

\begin{figure}
\epsfxsize=200pt
\centerline{\epsffile{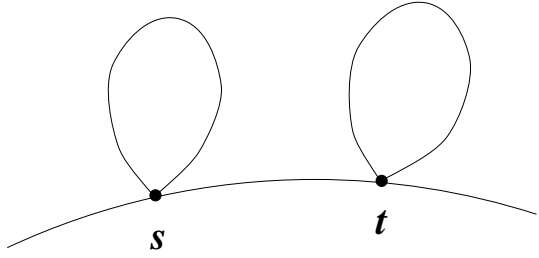}}
\caption{Parameter dependent area derivatives that act on different 
points of the main loop commute since it does not matter in which order
the infinitesimal loops are inserted.}
\end{figure}

\begin{figure}
\epsfxsize=200pt
\centerline{\epsffile{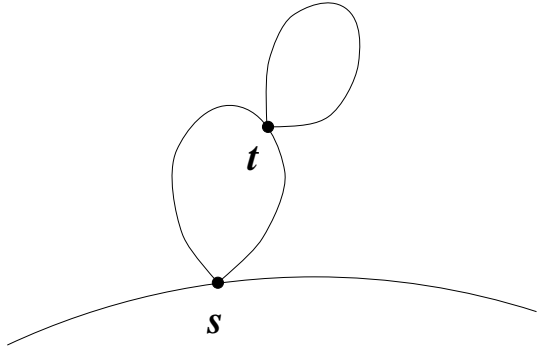}}
\caption{Two area derivatives at the same point do not
commute since there is a non-vanishing contribution when one of the
small loops is inserted onto the other.}
\end{figure}

To motivate the relation between the parameter dependent area
derivative and the ordinary functional derivative with respect to a
loop, compare the transfer relation (\ref{transfer}) with the
Mandelstam equation,
\beq
   \Delta_{ab}(s) \tr U_\eta = F^i_{ab}(\eta(s)) T^i_s \tr U_\eta.
\label{mandeleq}
\eeq
All that is missing is the contraction with a tangent vector, and 
one can show that for smooth loops,
\beq
   \frac{\d}{\d \eta^a(s)} = \dot\eta^b(s) \Delta_{ab}(s).
\label{delDels}
\eeq
And as we already pointed out, the commutator of two functional
derivatives vanishes.

To summarize, the three types of derivatives that we consider are in
increasing order of generality the functional loop derivative, the
parameter dependent area derivative, and the path dependent area
derivative. They are related by (\ref{DelsDelx}) and
(\ref{delDels}). The basic commutators are the more complicated the
more general the derivative is, compare (\ref{deldel}),
(\ref{DelsDels}), and (\ref{DelxDelx}).

That suggests that using parameter dependent area derivatives already
offer a simplification over path dependent area derivatives, but using
functional loop derivatives is even simpler. In section 4, we
compute the commutators of the constraints. It turns out that most of
the calculations can be performed on the level of the functional loop
derivatives, but at one point we fall back onto parameter dependent
area derivatives. While it is not proven that it is not possible to
perform the calculation exclusively in terms of functional
derivatives, it is a convenient approach. 

As already emphasized, however, the Hamiltonian constraint is not just
a derivative operator, but contains a rerouting as an additional
complication, which we address in section 3.2.


\subsection{Commutators of loop derivatives on piecewise smooth loops}

The loop representation allows continuous, piecewise smooth loops,
which means there may be kinks, $\dot\eta^a(s^-) \neq \dot\eta^a(s^+)$.
The area derivative is obviously well-defined at kinks, e.g.\
(\ref{mandeleq}) (that is, kinks do not prevent a loop functional from
being area-differentiable). As evident from $\frac{\d}{\d \eta^a(s)} =
\dot\eta^b(s) \Delta_{ab}(s)$, the functional loop derivative is 
ill-defined at kinks where $\dot\eta^b(s)$ does not exist.

For loops with kinks, integration around the loop as in $\ints
\dot\eta^a(s)$ is defined in terms of left- and right-handed
limits the same way we resolved the ambiguity in $T_s^i T_t^j$ for $s
= t$. Even if one assumes that the loop argument of $\psi[\eta]$ is
smooth, action by the metric or the Hamiltonian constraint introduces
kinks at intersections. Recall that 
\beq
[ \frac{\d}{\d \eta^a(t)}, T_t^j ] := 
\frac{\d}{\d \eta^a(t^-)} T_t^j - \frac{\d}{\d \eta^a(t^+)} T_t^j,
\label{comdel}
\eeq
and hence the loop derivative is kept away from the kinks introduced
by the rerouting. 

But notice that there are now two limits involved.
Suppose that there is a kink at $t_0$. In the limit
that the regulators are removed, one of the terms in
the Hamiltonian arises for $t$ in (\ref{hamil}) close
to $t_0$. We impose that the limit in the $t^\pm$ has to be
taken {\it before} the limit of $t\rightarrow t_0$. That means that
always either $t^-<t<t^+<t_0$ or $t_0<t^-<t<t^+$.
This definition of the integrals in the metric and the Hamiltonian
constraint together with the point-splitting regularization, which led
to our prescription for the positioning of the loop derivative, insure
that the constraints are unambiguously defined for kinks.

We therefore have for the commutators between reroutings and
functional loop derivatives for integrations of the type that appears
in the constraints (i.e.\ with the appropriate left- and right-sided
splits in the range of integration),
\beqa
   && \ints\intt f(s,t) [T_s^i,T_t^j] = 0,
\label{cc1}
\\
   && \ints\intt\intu g(s,t,u) [\frac{\d}{\d \eta^a(u)}, 
      T_s^j [\frac{\d}{\d \eta^a(t)}, T_t^j  ]] = 0,
\label{cc2}
\\
   && \ints\intt\intu\intv h(s,t,u,v) 
     [T_s^j [\frac{\d}{\d \eta^a(t)}, T_t^j  ],
      T_u^j [\frac{\d}{\d \eta^a(v)}, T_v^j  ] ] = 0,
\label{cc3}
\eeqa
for continuous functions $f$, $g$, and $h$ that maintain
reparametrization invariance. For example, (\ref{cc1})
follows from $[T_s^i,T_t^j] = \d_{st} \e^{ijk} T_s^k$ as long as
f(s,t) is assumed to be continuous. That is, there are contributions from 
the commutators defined by the left- and right-sided limits, but
the integral is zero as long as these contributions are finite and have
support only on sets of measure zero.
Of course, the reason why the commutators of the constraints are not
trivially zero because of (\ref{cc1})--(\ref{cc3}) is that 
functional differentiation can lead to distributional coefficients
$f$, $g$ or $h$.

Now we are ready to compute the constraint algebra.  In section 2 we
promised a demonstration that the unusual form of the Hamiltonian
constraint, that it involves only functional derivatives, reduces to
the standard form with area derivatives at kinks when the regulator is
removed. We do not actually remove the regulator in this paper, since
this would involve several case distinctions for smooth portions of
the loop, kinks and intersections, but as an important example
consider the Hamiltonian constraint for a loop with intersections.  
Suppose $\eta(s_0) = \eta(t_0)$ for $s_0 < t_0$. One of
the terms that arises due to the rerouting refers to
$\psi[\eta_{s_0t_0}]$, i.e.\ for a generic intersection there is a
kink at $s_0$, $t_0$. First notice that $\Delta_{ab}(s_0)
\psi[\eta_{s_0t_0}]$ is well defined, in fact,
\beq
      \Delta_{ab}(s_0) \psi[\eta_{s_0t_0}]
=     \Delta_{ab}(t_0) \psi[\eta_{s_0t_0}].
\label{Delatkink}
\eeq
Our definition of the Hamiltonian constraint gives rise to four terms
with a functional derivative given by the four orderings of $t_0$,
$t$, and $t^\pm$. For example, for $t_0<t^-<t<t+$,
\beq
   \lim_{t>t_0,t\rightarrow t_0}  \frac{\d}{\d \eta^a(t^-)}
   \psi[\eta_{s_0t_0}\eta_{t_0t^-}\eta_{t^-t}]
=
   \dot\eta^b(t_0^+) \Delta_{ab}(t_0) \psi[\eta_{s_0t_0}].
\eeq
The point is that in the limit that the regulator is removed we
recover the standard result (e.g.\ \cite{BrPu}) that in the generic
case, when the loops do not run smoothly through the intersection but
the tangentvectors $\dot\eta^a(s_0^-)$, $\dot\eta^a(s_0^+)$,
$\dot\eta^a(t_0^-)$, and $\dot\eta^a(t_0^+)$ are all different, there
are terms for which $\dot\eta^a(s)
\dot\eta^b(t) \Delta_{ab}(t)$ cannot be combined into loop derivatives
since the area derivative and tangent vectors are located on different
legs of the intersection.

The form of the Hamiltonian constraint (\ref{hamil}), and the
prescription for the order in which to take the two limits related to
reroutings is adopted precisely because it allows us to always keep the
area derivative and one tangent vector on the same leg so that they combine
to a functional loop derivative.


\section{The constraint algebra in the loop representation}

\subsection{Commutator of $D(v)$ with $D(w)$}

Given (\ref{deldel}), we immediately have that
\beqa
&& [ D(v), D(w) ] \psi[\eta]
\nonumber
\\
&=&
\ints \intt v^a(\eta(s)) \left(\frac{\d}{\d \eta^a(s)} w^b(\eta(t))\right)  
\frac{\d}{\d \eta^b(t)} \psi[\eta]  
\nonumber
\\
&& -
\ints \intt w^a(\eta(s)) \left(\frac{\d}{\d \eta^a(s)} v^b(\eta(t))\right)  
\frac{\d}{\d \eta^b(t)} \psi[\eta]  
\\
&=&
\ints \left( v^a(\eta(s)) \partial_a w^b(\eta(s))
- w^a(\eta(s)) \partial_a v^b(\eta(s)) \right)
\frac{\d}{\d \eta^b(s)} \psi[\eta]
\\
&=&
D({\cal L}_v w).
\eeqa
The same calculation can be performed after replacing $\d/\d\eta(s)$
by $\dot\eta^b(s)\Delta_{ab}(s)$, where the non-vanishing commutator
of the area derivatives cancels the contribution from the variation of
the tangent vectors. Much more involved is the direct use of path
dependent area derivatives
\cite{GaGaPu}, where the non-vanishing basic commutator cancels together
with the variation of the tangent vectors and the path dependence in
the area derivative when the Bianchi identity for area derivatives is used.
The basic structure in all three cases is that the variation of the smearing
vector fields gives rise to the relevant term in the commutator
without contribution from the other variations.

\subsection{Commutator of $D(v)$ with $H_{\e\e'}(N)$}

By definition of the constraints and (\ref{cc2}) we have
\beqa
&&
[ D(v), H_{\e\e'}(N) ]
\nonumber \\ &=&
\intx\ints\intt\intu N(x) 
v^a(\eta(u))
\frac{\d}{\d \eta^a(u)}
\left( f_\e(x,\eta(s)) f_{\e'}(x,\eta(t))  \dot\eta^b(s)\right)
\nonumber \\ && \quad
T_s^j [ \frac{\d}{\d \eta^b(t)} , T_t^j ]
\nonumber \\ && 
- \intx\ints\intt\intu N(x)
f_\e(x,\eta(s)) f_{\e'}(x,\eta(t)) \dot\eta^a(s) 
\nonumber \\ && \quad
T_s^j  [ \frac{\d}{\d \eta^a(t)} v^b(\eta(u)) , T_t^j ]
\frac{\d}{\d\eta^b(u)}.
\eeqa
The functional differentiation of the tangent vector gives
\beq
\intu  v^a(\eta(u)) \frac{\d}{\d\eta^a(u)} \dot\eta^b(s)
=
\frac{d}{ds} v^b(\eta(s)) 
=
\dot\eta^a(s)  \partial_a v^b(\eta(s)),
\eeq
while for the vector field,
\beq
\intu \frac{\d}{\d\eta^a(t)} v^b(\eta(u)) \frac{\d}{\d\eta^b(u)}
=
\partial_a v^b(\eta(t)) \frac{\d}{\d\eta^b(t)}.
\eeq
Since in the limit that the regulators are removed
$\eta(s)\simeq\eta(t)$, these two terms cancel in the commutator,
\beqa
&&
[ D(v), H_{\e\e'}(N) ] 
\nonumber \\ &\simeq&
\intx\ints\intt\intu N(x) 
v^a(\eta(u))
\frac{\d}{\d\eta^a(u)}
\left( f_\e(x,\eta(s)) f_{\e'}(x,\eta(t))\right)  \dot\eta^b(s)
\nonumber \\ && \quad
T_s^j [ \frac{\d}{\d \eta^b(t)} , T_t^j ].
\label{dh1}
\eeqa
For the functional differentiation of the regulators we obtain
\beqa
&&
\intx\intu N(x)
v^a(\eta(u))
\frac{\d}{\d\eta^a(u)}
\left( f_\e(x,\eta(s)) f_{\e'}(x,\eta(t))\right)
\nonumber \\ &=&
\intx N(x) ( 
- v^a(\eta(s))  f_{\e'}(x,\eta(t))  \partial_a f_\e(x,\eta(s))
- v^a(\eta(t)) f_\e(x,\eta(s)) \partial_a f_{\e'}(x,\eta(t))    )
\\ &=&
\intx (
v^a(\eta(s))  f_\e(x,\eta(s))  \partial_a (N(x) f_{\e'}(x,\eta(t)))
+ v^a(\eta(t)) f_{\e'}(x,\eta(t))  \partial_a (N(x) f_\e(x,\eta(s)))    )
\nonumber \\ &&
\mbox{} 
\\ &\simeq&
\intx (
v^a(x)  f_\e(x,\eta(s))  \partial_a (N(x) f_{\e'}(x,\eta(t)))
- \partial_a (v^a(x) f_{\e'}(x,\eta(t)) )  N(x) f_\e(x,\eta(s))  )
\\ &=&
\intx (v^a(x)\partial_aN(x) - N(x) \partial_a v^a(x)) 
f_\e(x,\eta(s)) f_{\e'}(x,\eta(t)).
\label{dh2}
\eeqa
We performed two partial integrations in order to remove the
derivative from the regulators, and we moved the vector field from the
loop to $x$. 
Note that in general $v^a(y)\partial_a\d^3(x-y) \neq
v^a(x)\partial_a\d^3(x-y)$, and that we first removed the partial
derivative from the regulator before concluding $v^a(\eta(s)) \simeq v^a(x)$.

Since $N$ is a scalar density of weight $-1$, ${\cal L}_vN=v^a\partial_aN 
- N\partial_av^a$, and inserting (\ref{dh2}) into
(\ref{dh1}) we obtain the operator version of the classical Poisson
bracket,
\beq 
[ D(v), H_{\e\e'}(N) ]  \simeq H_{\e\e'} ( {\cal L}_vN ).
\eeq


\subsection{Commutator of $H_{\d\d'}(M)$ with $H_{\e\e'}(N)$}

By definition of the constraints and (\ref{cc3}) we have
\beqa
&& 
[ H_{\d\d'}(M), H_{\e\e'}(N) ]
\nonumber \\ &=&
\intx\inty\ints\intt\intu\intv M(x) N(y)
f_\d(x,\eta(s)) f_{\d'}(x,\eta(t)) \dot\eta^a(s) T^j_s
\nonumber \\ && \quad
[ \frac{\d}{\d\eta^a(t)}
\left( f_\e(y,\eta(u)) f_{\e'}(y,\eta(v))  \dot\eta^b(u)\right) , T^j_t ]
T_u^k [ \frac{\d}{\d\eta^b(v)} , T_v^k ]
\nonumber \\ &&  -
(M\leftrightarrow N, \d\leftrightarrow\e, \d'\leftrightarrow\e').
\label{hamhaminit}
\eeqa
Functional differentiation of the tangent vector gives
\beqa
\intu f_\e(y,\eta(u)) T_u^k \frac{\d}{\d\eta^a(t)} \dot\eta^b(u)
&=& 
- \frac{d}{dt} (f_\e(y,\eta(t)) T_t^k) \d^b_a
\\ &=&
\dot\eta^c(t) \partial_c f_\e(y,\eta(t)) T_t^k \d^b_a
- f_\e(y,\eta(t)) \frac{d}{dt} 
T_t^k \d^b_a.
\eeqa
The second term does not contribute to the commutator since
\beqa
&&
M(x) N(y) 
f_\d(x,\eta(s)) f_{\d'}(x,\eta(t)) 
f_\e(y,\eta(t)) f_{\e'}(y,\eta(v))
\nonumber \\ &&
-
N(x) M(y) 
f_\e(x,\eta(s)) f_{\e'}(x,\eta(t)) 
f_\d(y,\eta(t)) f_{\d'}(y,\eta(v))
\simeq 0.
\label{mnnm}
\eeqa
Therefore, the functional differentiation of the tangent vector in the
commutator gives  
\beqa
&&
\intx\inty\ints\intt\intu
M(x) N(y) 
f_\d(x,\eta(s)) f_{\d'}(x,\eta(t)) 
\partial_a f_\e(y,\eta(t)) f_{\e'}(y,\eta(u))
\nonumber \\ && \quad
\dot\eta^b(s) \dot\eta^a(t)
T_s^j [T_t^k,T_t^j] [\frac{\d}{\d\eta^b(u)},T_u^k]
\nonumber \\ &&
 -
(M\leftrightarrow N, \d\leftrightarrow\e, \d'\leftrightarrow\e').
\eeqa
To remove the partial derivative from the regulator, we perform a
partial integration in $y$, and with approximations similar to
(\ref{mnnm}), 
\beqa
&&
\intx\inty M(x) N(y) 
f_\d(x,\eta(s)) f_{\d'}(x,\eta(t)) 
\partial_a f_\e(y,\eta(t)) f_{\e'}(y,\eta(u))
\nonumber \\ && 
-
\intx\inty N(x) M(y) 
f_\e(x,\eta(s)) f_{\e'}(x,\eta(t)) 
\partial_a f_\d(y,\eta(t)) f_{\d'}(y,\eta(u))
\nonumber \\ &\simeq& 
- \intx\inty \w_a(x) 
f_\d(x,\eta(s)) f_{\d'}(x,\eta(t)) 
f_\e(y,\eta(t)) f_{\e'}(y,\eta(u))
\\ &\simeq&
- \w_a(\eta(u)) f_\d(\eta(u),\eta(s)) f_\e(\eta(u),\eta(t)),
\label{regrem}
\eeqa
where
\beq
\w_a(x) 
=
M(x) \partial_a N(x) - N(x) \partial_a M(x).
\label{omega}
\eeq
($\w_a$ is a covector density of weight -2, and $\w_a f_\d f_\e$ has
weight 0.)  
In (\ref{regrem}) we integrate over $x$ and $y$ keeping terms to
leading order. Note that already in the definition
(\ref{hamil}) of the Hamiltonian constraint one of the regulators can
be removed without any renormalization, say $f_{\e'}(x,\eta(t))
\rightarrow \d^3(x,\eta(t))$. But after integrating over $x$, it is
then not quite clear how to perform the necessary partial integrations
in the commutator algebra.

Functional differentiation of the regulators in the commutator
(\ref{hamhaminit}) leads directly to partial derivatives of
regulators, which can be removed analogously, and we obtain for the
full commutator (after appropriate renaming of integration variables
and contracted indices)
\beqa
&&
[ H_{\d\d'}(M), H_{\e\e'}(N) ]
\nonumber \\ &\simeq&
- \ints\intt\intu 
\w_a(\eta(u)) f_\d(\eta(u),\eta(s)) f_\e(\eta(u),\eta(t))
\dot\eta^a(s) \dot\eta^b(t)
\nonumber \\ && \quad
\left(
[ T_s^k, T_s^j] T_t^j [\frac{\d}{\d\eta^b(u)},T_u^k]
+
T_s^j [ T_t^k, T_t^j] [\frac{\d}{\d\eta^b(u)},T_u^k]
+
T_s^j T_t^k [ [ \frac{\d}{\d\eta^b(u)} , T_u^k] , T_u^j]
\right).
\label{hamhamalmost}
\eeqa
A priori it is not obvious how the right-hand side of the classical
Poisson bracket (\ref{hh}), $D(g^{ab}\w_b)$, should be represented
in the loop representation because of the operator product of the
metric with the generator of diffeomorphisms. Comparing the above with
the loop operators for the diffeomorphism constraint (\ref{diffeo})
and the metric (\ref{metric}), a natural guess is that the reroutings
in (\ref{hamhamalmost}) combine to $T_s^j T_t^j
\frac{\d}{\d\eta^b(u)}$. This is indeed the case.

The first two terms in (\ref{hamhamalmost}) cancel since with
(\ref{com2eps}),
\beqa
&&
[ T_s^k, T_s^j] T_t^j [\frac{\d}{\d\eta^b(u)},T_u^k]
+
T_s^j [ T_t^k, T_t^j] [\frac{\d}{\d\eta^b(u)},T_u^k]
\nonumber \\ &=&
[\frac{\d}{\d\eta^b(u)}, 
\e^{ikj} T_s^i T_t^j T^k_u
+
\e^{ikj} T_s^j T_t^i T^k_u] 
\\ &=& 0.
\eeqa
Note that the cancellation does not depend on whether
$\frac{\d}{\d\eta^b(u)}$ is located at $u^-$ or $u^+$.

However, the one remaining term cannot be simplified using the $SU(2)$
commutator for the insertion operators, 
because resolving the order of the insertions puts the derivative
between the insertion operators, 
\beq
[ [ \frac{\d}{\d\eta^b(u)} , T_u^k] , T_u^j] 
=
\frac{\d}{\d\eta^b(u^-)} T_u^k T_{u^+}^j
-
T_u^k \frac{\d}{\d\eta^b(u^+)} T_{u^{++}}^j
-
T_{u^{--}}^j \frac{\d}{\d\eta^b(u^-)} T_u^k
+
T_{u^-}^j T_u^k \frac{\d}{\d\eta^b(u^+)}.
\eeq
Of course, there are other identities to work with, but let us first
consider the situation in the connection representation, i.e. 
$\psi[\eta] = \tr U_\eta$ and 
$\frac{\d}{\d\eta^b(u)} = \dot\eta^b(u) F^i_{ab}(\eta(u)) T^i_u$.
Then the rerouting simplifies according to
\beq
	T_s^j T_t^k [[ T_u^i, T_u^k], T_u^j] 
=
        T_s^j T_t^k \e^{ikl} \e^{ljm} T^m_u
=
        T_s^i T_t^j T_u^j - T_s^j T_t^j T_u^i.
\eeq
The important observation is that this result cannot be 
transformed back to a functional derivative and some reroutings in the
loop representation because in the first term $F^i_{ab}(\eta(u))$ is
contracted into an insertion at $s$, not $u$. In the limit that the
regulators are removed in (\ref{hamhamalmost}), we have
$\eta(u)\simeq\eta(s)$,
and
\beq
	T_s^j T_t^k [ [ \frac{\d}{\d\eta^b(u)} , T_u^k] , T_u^j] 
=
	\dot\eta^c(u) \Delta_{bc}(s) T^j_t T^j_u - 
        T_s^j T_t^j \frac{\d}{\d\eta^b(u)}.
\label{combi}
\eeq
As already discussed in section 3.2, the reroutings in the
Hamiltonian may split a functional loop derivative into a tangent
vector and an area derivative at different parameters. 

While by definition of the loop representation its states have to
satisfy the loop transforms of all the $SU(2)$ identities of the
connection representation, i.e. in particular (\ref{combi}), let us
also indicate how (\ref{combi}) can be derived using the $SU(2)$
identities in the loop representation without resorting to the loop
transform. For definiteness, consider the case $s<t<u$. The rerouting
operations that appear in (\ref{combi}) are
\beqa 
	T_s^j T_t^j \psi[\eta] &=& \fof\psi[\eta] - 
       \fot \psi[\eta_{st}\cup\eta_{ts}], 
\\
        T_{s^+}^j T_{t^+}^j \psi[\eta_{st}\cup\eta_{ts}]
&=&     \fof \psi[\eta_{st}\cup\eta_{ts}] - \fot \psi[\eta].
\eeqa
Marking the area derivative by the insertion of a small loop $\c$, the
claimed relation (\ref{combi}) becomes
\beqa
&&
	\psi[\eta_{st}\cup\eta_{us}\cup\eta_{tu}\c]
-
        \psi[\eta_{tu}\cup\eta_{us}\cup\eta_{st}\c]
+ 
        \psi[\eta_{us}\eta_{tu}\eta_{st}\c]
-
        \psi[\eta_{tu}\eta_{st}\eta_{us}\c]
\nonumber \\ &=&
        2 \psi[\eta_{st}\cup\eta_{us}\eta_{tu}\c]
-
        2 \psi[\eta_{tu}\cup\eta_{st}\eta_{us}\c]
+
        \psi[\eta_{st}\eta_{tu}\eta_{us}\c]
-
        \psi[\eta_{us}\eta_{st}\eta_{tu}\c].
\eeqa
A simple systematic method to proceed, without introducing the
complication of inverted loops as in the standard spinor identity, can
be found in \cite{BrPu:sol}. First, use (43) of \cite{BrPu:sol}
(there is a factor of two missing on the left-hand side) to resolve
any $\psi[\a\b\c\d]$ into a sum of loop states depending on multiloops
where a single loop contains at most three of the four loops $\a$,
$\b$, $\c$, and $\d$. Then use (40) of \cite{BrPu:sol} to rewrite
any $\psi[\b\a\c]$ in some preferred order $\psi[\a\b\c]$ plus a sum
of loop states depending only on multiloops where a single loop
contains at most two of the three loops.  Then our claim becomes
\beq
	\psi[\c\cup\eta_{tu}\cup\eta_{us}\eta_{st}]
-
        \psi[\c\cup\eta_{st}\cup\eta_{tu}\eta_{us}]
=
0,
\eeq
which holds because $\c$ is inserted on the trivial loop, and the area
derivative of a trivial loop dependence vanishes. Note that we used
(\ref{Delatkink}) for the area derivative at a kink several times.

Given (\ref{combi}), the final observation is that in the commutator
$\dot\eta^b(t)\dot\eta^c(u)\Delta_{bc}(s)$ is anti-symmetric under
exchange of $t$ and $u$, while the remainder of the integral is
symmetric under exchange of $t$ and $u$. Therefore, the term in the
integrand that does not reduce to a functional loop derivative
vanishes under the integral. Hence we have shown that
\beq
[ H_{\d\d'}(M), H_{\e\e'}(N) ] 
\simeq
\intu \w_a(\eta(u)) g^{ab}_{\d\e}(\eta(u))
\frac{\d}{\d\eta^b(u)}.
\eeq
The result represents one of the possible factor orderings of the
operator version of ${D(\w_a g^{ab})}$, which appears in the classical
Poisson bracket.

\section{Conclusion}

We have confirmed the result of \cite{GaGaPu} that the constraint
algebra of 3+1 quantum gravity in the loop representation formally
closes without anomalies. There are two reasons for the comparative
simplicity of our calculation. We were able to cast the Hamiltonian
constraint into a form involving only functional loop derivatives
instead of area derivatives, and we found a simple way to separate the
rerouting operations from the other parts of the calculation.

One direction for further research is to analyze the point-splitting
regularization in more detail, in particular to analyze the next to
leading order terms \cite{Bo}. This appears to be necessary if one
decides to take the point-splitting regularization seriously since one
cannot remove the regulators in the definition of the constraints
before computing their algebra without running into inconsistencies
related to the background dependence of the regularization (also see
the comment following (\ref{omega})). Turning
this observation around, we avoid anomalous background dependent terms
in the constraint algebra by postponing the removal of the regulators
until the algebra is computed.

A related point is that if we remove the regulator, the Hamiltonian
constraint consists of discrete sums over kinks and intersections (no
integrals involved) and integrals along the loops for the acceleration
terms \cite{Bl,BrPu,Bo}. The acceleration terms in a sense spoil the
simple picture that the Hamiltonian constraint only acts on
intersections, which are invariant under diffeomorphisms, although the
background is present as an angle dependence, and the acceleration
terms depend on the background since they contain second derivatives
of the loop, $\ddot\eta^a(s)$.  Sometimes one would like to argue away
the acceleration terms, but notice that in order to obtain the integral
on the right-hand side of the commutator of two Hamiltonians,
(\ref{hh}), and as is also apparent from
section 4.3, it does not suffice to just consider the discrete sums
corresponding to kinks and intersections. This may still be consistent
with a diffeomorphism invariant scheme like that of
\cite{RoSm:prl}, in which acceleration terms do not appear, since, as
we discussed in the introduction, (\ref{hh}) is no longer relevant.

As a final remark, note that the rigorous framework based on
diffeomorphism invariant measures \cite{AsLeMaMoTh,AsLe,Ba} is well
adapted to the generators of diffeomorphisms, but has problems with
the area derivatives appearing in the Hamiltonian constraint.
Therefore it may be worthwhile to examine our form of the Hamiltonian
constraint (\ref{hamil}) in that setting.

It is a pleasure to thank Abhay Ashtekar, John Baez, Roumen Borissov,
Renate Loll and Carlo Rovelli for helpful discussions.

\newcommand{\bib}[1]{\bibitem{#1}}

\newcommand{\apny}[1]{{\em Ann.\ Phys.\ (N.Y.) }{\bf #1}}
\newcommand{\cjm}[1]{{\em Canadian\ J.\ Math.\ }{\bf #1}}
\newcommand{\cmp}[1]{{\em Commun.\ Math.\ Phys.\ }{\bf #1}}
\newcommand{\cqg}[1]{{\em Class.\ Quan.\ Grav.\ }{\bf #1}}
\newcommand{\grg}[1]{{\em Gen.\ Rel.\ Grav.\ }{\bf #1}}
\newcommand{\jgp}[1]{{\em J. Geom.\ Phys.\ }{\bf #1}}
\newcommand{\ijmp}[1]{{\em Int.\ J. Mod.\ Phys.\ }{\bf #1}}
\newcommand{\jmp}[1]{{\em J. Math.\ Phys.\ }{\bf #1}}
\newcommand{\mpl}[1]{{\em Mod.\ Phys.\ Lett.\ }{\bf #1}}
\newcommand{\np}[1]{{\em Nucl.\ Phys.\ }{\bf #1}}
\newcommand{\pl}[1]{{\em Phys.\ Lett.\ }{\bf #1}}
\newcommand{\pr}[1]{{\em Phys.\ Rev.\ }{\bf #1}}
\newcommand{\prl}[1]{{\em Phys.\ Rev.\ Lett.\ }{\bf #1}}
\newcommand{\bb}{B. Br\"ugmann}

\end{document}